\documentstyle[preprint,aps,epsfig]{revtex}

\begin{document}
\draft
\title{Spin Injection from Ferromagnetic Metals into Gallium Nitride}

\author{C. J. Hill, X. Cartoix\`{a}, R. A. Beach}
\address{T. J. Watson, Sr., Laboratory of Applied Physics,
California Institute of Technology,
Pasadena, California 91125 USA}
\author{D. L. Smith}
\address{Los Alamos National Laboratory,Los Alamos, NM 87545 USA}
\author{T. C. McGill}
\address{T. J. Watson, Sr., Laboratory of Applied Physics,
California Institute of Technology,
Pasadena, California 91125 USA}
\date{\today}

\maketitle
\begin{abstract}
The injection of spin polarized electrons from ferromagnetic metals (Fe and Co) into gallium nitride (GaN) via scanning tunneling microscopy (STM) is demonstrated.  Electrons from STM tips are injected into the semiconductor.  Net circular polarization of the emitted light is observed, which changes sign on reversal of the magnetization of the tip. The polarization is found to be in qualitative agreement with that expected from considerations based on the splitting of the valence bands due to spin-orbit coupling and the crystal field splitting corresponding to the wurtzite structure, and the magnitude of the spin polarization from the ferromagnetic metal.  We find a lower bound for the spin injection efficiency of 25\%, corresponding to a net spin polarization in the semiconductor of 10\%. This is the largest reported value for a room temperature measurement of spin injection into semiconductors in air.
  
\end{abstract}

The focus of the new discipline of spintronics is the control and utilization of spin in electronic devices~\cite{Wolf2000,Fitzgerald2000,Studt1999}.  A central issue in spintronics is whether one can inject spin polarized currents into electronically useful materials.  Historically, the most useful electronic materials are the semiconductors and the most convenient spin polarized sources appear to be the ferromagnetic metals.  The attempts to inject spin polarized electrons from ferromagnetic metals into semiconductors have been the subject of a great deal of effort but to date have led to inconclusive results \cite{HammarBennettYang1999,MonzonTangRoukes2000,Vanwees2000,HammarBennettYang2000:prl}.  In the last year success has been achieved by electrically injecting from a spin polarized semiconductor into a nonmagnetic semiconductor \cite{FiederlingKeimReuscher1999,OhnoYoungBeschoten1999,Jonker2000}.
The polarization of emitted photons is used as a measure of the net spin polarization.  These experiments all
require low temperatures and high magnetic fields to produce nearly 100\% polarization of the electron spins in the spin polarized semiconductor contact.
Alvarado and coworkers \cite{AlvaradoRenaud1992,Alvarado1993} have carried out experiments using a scanning tunneling microscope (STM) tip made of Ni to inject electrons into p-GaAs that has been cleaved in an ultra-high vacuum (UHV) environment.   This approach is not particularly useful for making devices because of the requirement of a UHV environment. The success of the spin-polarized-semiconductor/nonmagnetic-semiconductor injection experiments as well
as the failure to inject spin polarized currents from a metal contact to a semiconductor have been accounted for in an irreversible thermodynamics theory~\cite{SchmidtFerrandMolenkamp2000}.

Here we present the results of spin injection by scanning tunneling microscopy (STM) into p-GaN layers in air. The injected spin polarization is determined by measuring the net circular polarization of the emitted light.  We find that the light has a net circular polarization
that changes sign when the magnetization direction of the ferromagnetic STM tip is changed.  

The samples were grown in a Superior Vacuum Technology Associates BLT N35 molecular beam epitaxy machine. A 0.6 $\mu$m GaN p-type layer
doped at about $10^{16}$ $\text{cm}^{-3}$ with Mg was grown on a sapphire substrate with a 1 $\mu$m GaN buffer layer.  Gold was deposited on top
of the sample and then patterned using standard lift-off techniques to expose 500 $\mu$m holes in the Au layer for probing. The experimental apparatus (see Fig.~\ref{fig:ExpSetup}) consisted of a Digital Instruments Multimode scanning tunneling microscope probe brought in close proximity to the p-GaN layer.  Electrons are injected into the p-GaN layer.  The light emitted from the GaN layer is analyzed for circular polarization using a $\lambda$/4 waveplate and a linear polarizer alternately positioned to select right or left circular polarization.  This arrangement insures that the polarization will not be altered by any of the remaining optical elements.  The light is collected by a microscope objective lens and passed thru a light pipe to a Hamamatsu model 943-02 photomultipler tube for detection.  To produce a luminescence signal about a hundred times higher than the dark counts in the photomultipler tube, the injection current was typically 5 to 10 $\mu$A.  To achieve these high currents the STM was operated in an I-V mode where the feedback loop maintains the tip at close proximity to the surface and then the applied voltage between the gold layer and the tip is increased to 9.5 V for the duration of data collection. We believe this non-standard feedback loop was the cause of a substantial drift with time displayed by the data.

The major experimental results of this paper are shown in Figs.~\ref{fig:DataFe} and \ref{fig:DataCo} for Fe and Co respectively.  The two panels in each figure show the raw data for magnetizations parallel and antiparallel to the tip axis. The data points represent photon counts over 30 second intervals. The linear polarizer was rotated by 90$^{\circ}$ between subsequent points to distinguish left and right circularly polarized light. This procedure was used to circumvent the effects of long term drift in the photon count rate. The figures show clearly that there is a difference between the photon count rates for the two circular polarizations. In the top panel, where the magnetization is parallel to the tip axis, right circularly polarized (RCP) light predominates. In the lower panel, where the magnetization is antiparallel to the tip axis, the trend reverses and left circularly polarized (LCP) light predominates.  Systematic changes in the number of photon counts with the orientation of the linear polarizer demonstrate that one polarization dominates and that the dominant polarization changes with the direction of magnetization of the tip.

These results cannot be explained by local field effects from the magnetic tip polarizing the electrons in the GaN. A simple estimate based on the magnetic field produced by a ferromagnetic plate shows that the maximum field in this worst-case scenario would be less than 2 T, while the necessary field to produce a splitting $ g^{*} \mu_{B} B \sim kT$ at room temperature would be over 200 T, where $g^{*}$ is the effective $g$-factor for GaN, $\mu_{B}$ is the Bohr magneton, $B$ is the magnetic induction due to the tip, $k$ is the Boltzmann constant and $T$ is the temperature.  This experiment also does not suffer from local Hall effects which can give spurious signals in electrical measurements of spin injection~\cite{MonzonTangRoukes2000}.

Figure~\ref{fig:ExpPolar} is a summary of the observed polarization effects.  The experimental value for the optical polarization, defined as $(RCP-LCP)/(RCP+LCP)$ is between  2\%\ and 4\%\ for the two metals. The observation that the magnitude of the effect is similar for Fe and Co is in qualitative agreement with the measurements of Soulen et al. \cite{SoulenOsofskyNadgorny1999}, who used Andreev reflection to find that Co and Fe have roughly the same amount of spin polarization.  The sign of the dominant helicity (using the convention in optics that a RCP photon has negative helicity) is in agreement with majority spins in Fe and Co dominating electrical conduction and recombining predominantly with states in the GaN heavy hole band. 

To correlate the observed optical polarization with the spin polarization of the electrons injected by the metallic tip one must consider several material dependent factors.
The simplest is the fact that there is a tendency for the contribution from various interband transitions to cancel.  The conduction band and valence band structure in GaN are shown in Fig.~\ref{fig:bandplot}.  The conduction band edge is 
$\Gamma_{7c}$ and the valence band is split into three subbands $\Gamma_{9v}$ (heavy hole), $\Gamma_{7v}^{SO}$ (light hole) and $\Gamma_{7v}^{CR}$ (crystal split) \cite{YeoChongLi1998}.
For a single spin in the conduction band, the transitions from the $\Gamma_{7c}$ conduction band to the  $\Gamma_{9v}^{HH}$have one direction of circular polarization, while transitions to the $\Gamma_{7v}^{LH}$ and $\Gamma_{7v}^{CR}$ subbands have the opposite direction.  Since the amount of the splittings is comparable to $kT$, $k$ being the Boltzmann constant and $T$ the temperature, the probability of transitions to the $\Gamma_{7}$'s taking place will not be greatly diminished with respect to transitions to the $\Gamma_{9}$, and the net polarization will be reduced.  A simple use of the Wigner-Eckardt theorem with the Clebsch-Gordan coefficients corresponding to the wurtzite point group $C_{6v}$ \cite{Lax1974,KosterDimmockWheeler1963} yields the following expression for the net polarization for the case where the spins are pointing along the $z$ axis and the light is emitted along the $z$ axis as well
\begin{equation}
{NP}={NS}\times \frac {1-|a|^{2} \frac{m^{j}_{LH\perp} \sqrt{m^{j}_{LH\parallel}}} {m^{j}_{HH\perp} \sqrt{m^{j}_{HH\parallel}}} \exp(-E_1/{kT}) - |b|^{2} \frac{m^{j}_{CR\perp} \sqrt{m^{j}_{CR\parallel}}} {m^{j}_{HH\perp} \sqrt{m^{j}_{HH\parallel}}} \exp(-E_2/{kT})} {1+|a|^{2} \frac{m^{j}_{LH\perp} \sqrt{m^{j}_{LH\parallel}}} {m^{j}_{HH\perp} \sqrt{m^{j}_{HH\parallel}}} \exp(-E_1/{kT}) + |b|^{2} \frac{m^{j}_{CR\perp} \sqrt{m^{j}_{CR\parallel}}} {m^{j}_{HH\perp} \sqrt{m^{j}_{HH\parallel}}} \exp(-E_2/{kT})} 
\label{eq:Polarization}
\end{equation}
where a and b satisfy $1-|a|^{2}-|b|^{2}=0$ and are defined in Ref.~\cite{ChuangChang1996}, $E_1$ and $E_2$ are the energy splittings in the valence band as defined in Fig.~\ref{fig:bandplot}. The symbol $NP$ is the net polarization of the light.  $NS$ is the net spin polarization of the injected electrons, defined as \hbox{$(n_{\uparrow}-n_{\downarrow})/(n_{\uparrow}+n_{\downarrow})$} where $n_{\uparrow}$ ($n_{\downarrow}$) is the electron density with spin up (down) along the $z$ axis. The symbol $m^{j}_{HH\perp}$ ($m^{j}_{HH\parallel}$) is the joint effective mass for the conduction and the heavy hole band along the $z$ axis (in the $x-y$ plane), defined as
\begin{equation}
\frac {1}{m^{j}_{HH\perp}} = \frac {1}{m^{*}_{c\perp}} + \frac {1}{m^{*}_{HH\perp}}
\end{equation}
where $m^{*}_{c\perp}$ ($m^{*}_{HH\perp}$) is the effective mass for electrons in the conduction band (holes in the heavy hole band) along the $z$ axis. The rest of the symbols are defined analogously.

A limit on the ratio can be obtained by introducing the numerical values for the effective masses \cite{YeoChongLi1998} and the a and b coefficients \cite{ChuangChang1996}, obtaining
\begin{equation}
 {\text{Net Light Polarization}\over \text{Net Polarization of Injected Electrons}} \approx {0.31} 
\end{equation}
at room temperature.  Hence the measured ratio is reduced by the two competing transitions. Assuming  $NS=40\%$ for Fe and Co \cite{SoulenOsofskyNadgorny1999}, an upper bound for $NP<12\%$ is obtained.  Another factor is that some of the emitted light is below the band edge.  These 
transitions may not yield light with a definite polarity \cite{Jonker2000}, thus generally reducing the measured polarization. Further reduction can come from spin randomizing events taking place before the radiative recombination occurs.

We can deduce a minimum value for the efficiency of the spin injection process
\begin{equation}
\eta = \frac {\text{Spin Polarization in the Semiconductor}} {\text{Spin Polarization in the Ferromagnet}}.
\end{equation}
We find that $\eta$ is at least  25\%.

The observation of electrical spin injection from a ferromagnetic 
STM tip into a semiconductor reported here is consistent with the theoretical
results of Smith and Silver\cite{SmithSilver2000}.  The vacuum tunneling barrier that occurs
in the STM configuration provides a spin dependent interface resistance that,
when tunneling from a ferromagnetic tip, produces spin injection.

In summary, we observe the largest room temperature spin injection efficiency from a ferromagnetic metal into a semiconductor in air.  We have interpreted the data in terms of interband transitions and selection rules obtained by means of the Wigner-Eckardt theorem and the Clebsch-Gordan coefficients for point groups, and found it to be consistent with electrical conduction by the majority spins. We have also shown that local field effects are too small to produce any substantial spin splitting leading to an unwanted signal.  We have calculated a minimum spin injection efficiency of 25\%, and shown that our data is consistent with the view that majority spins from Fe and Co are injected into the GaN conduction band and recombine predominantly with heavy hole valence band states. 

\begin{acknowledgments}
We gratefully acknowledge J. O. McCaldin who has guided us in our search
for materials to gallium nitride. We wish to thank A. T. Hunter for his advice on optical detection techniques.  We would like to acknowledge technical advice from P. M. Bridger and G. S. Picus. B.T. Jonker of the Naval Research Laboratories pointed out the possibility of below band gap radiation not being polarized. We wish to acknowledge the support of the Defense Advanced Research Projects under the direction of S. A. Wolf and
monitored by the L. R. Cooper of the Office of Naval Research.
\end{acknowledgments}

\bibliography{prl1}
\bibliographystyle{prsty}


\begin{figure}
\end{figure}

\begin{itemize}
\item[Figure 1]Schematic of the experimental arrangement for measuring the spin polarization. The polarization analysis is carried out by the combination of the $\lambda$/4 waveplate and the linear polarizer before any of the other optical components.

\item[Figure 2]The time dependent experimental data for iron.  The data shows the results for a sequence of measurements with the polarization analyzer rotated periodically to distinguish left and right circularly polarized light.  The two curves are for the two orientations of the magnetization in the metallic tip.

\item[Figure 3]The time dependent experimental data for cobalt.  The data shows the results for a sequence of measurements with the polarization analyzer rotated periodically to distinguish left and right circularly polarized light.  The two curves are for the two orientations of the magnetization in the metallic tip.

\item[Figure 4]The results of the optical polarization measurements.  The measured degree of polarization is shown for Fe and Co for both tip magnetizations.

\item[Figure 5]Possible transitions from the conduction band into the valence band with polarization resolution. The short thick arrows represent the electron spin. For a given electron spin, transitions into the $\Gamma_{9}$ subband yield photons with opposite polarization than transitions into $\Gamma_{7}$ subbands.

\end{itemize}
\pagebreak
\begin{figure}
\centering
\epsfig{file=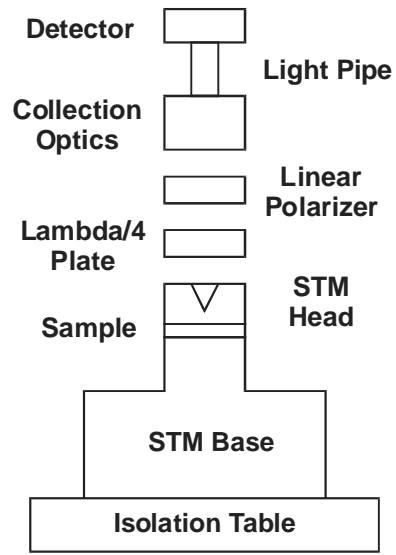,width=2in,clip=}
\vspace*{\fill}
\caption{{\large \bf C. J. Hill,  Physical Review Letters }}
\label{fig:ExpSetup}
\end{figure}
\clearpage

\begin{figure}
\centering
\epsfig{file=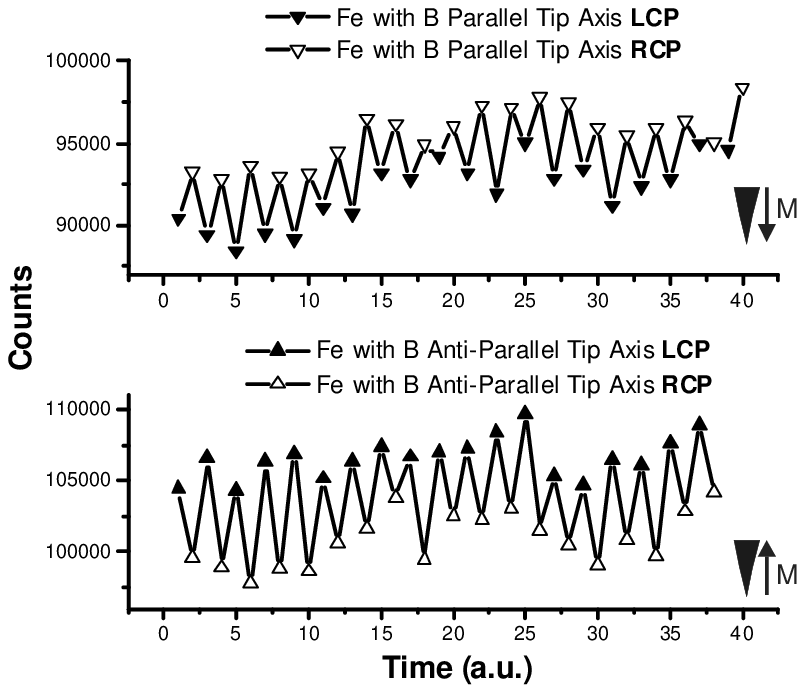,width=3.5in,clip=}
\vspace*{\fill}
\caption{{\large \bf C. J. Hill,  Physical Review Letters }}
\label{fig:DataFe}
\end{figure}
\clearpage

\begin{figure}
\centering
\epsfig{file=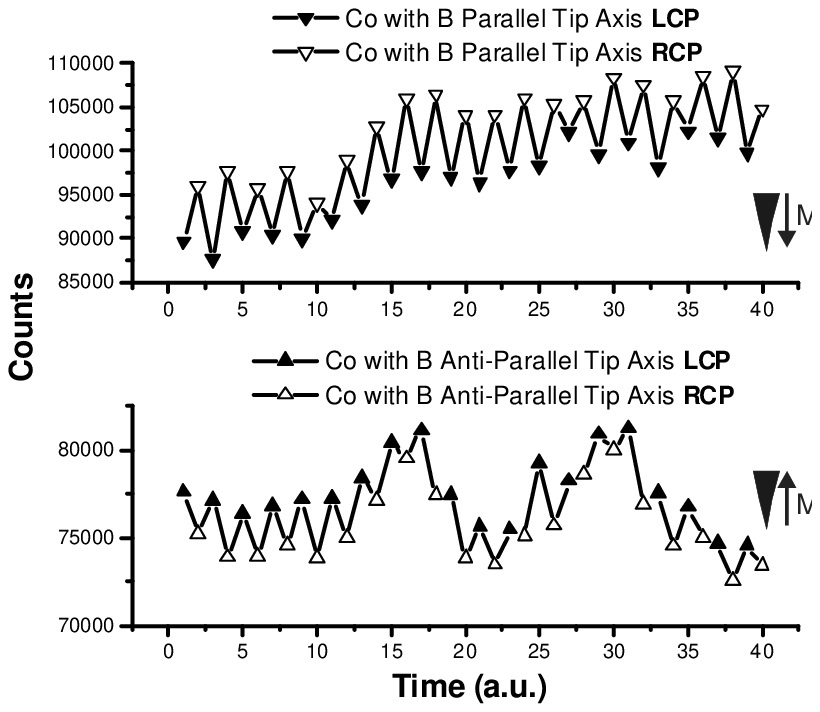,width=3.5in,clip=}
\vspace*{\fill}
\caption{{\large \bf C. J. Hill,  Physical Review Letters }}
\label{fig:DataCo}
\end{figure}
\clearpage

\begin{figure}
\centering
\epsfig{file=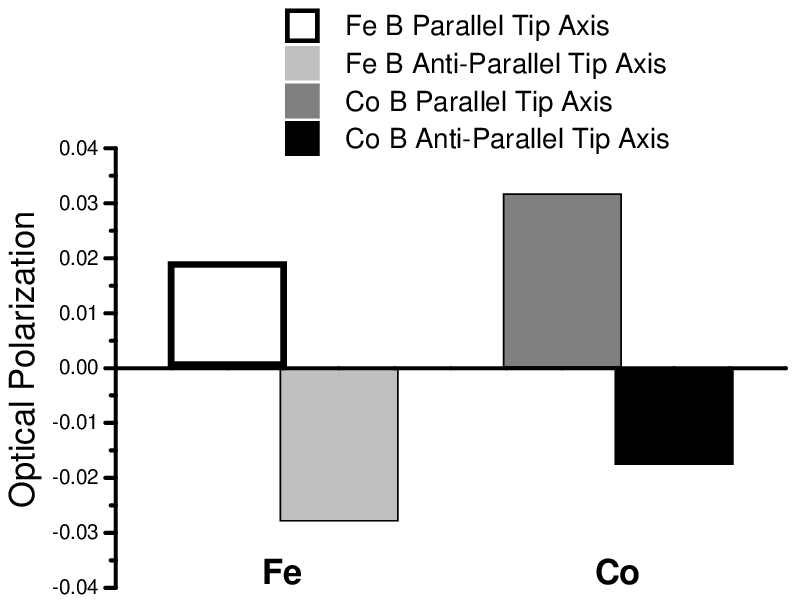,width=3.5in,clip=}
\vspace*{\fill}
\caption{{\large \bf C. J. Hill,  Physical Review Letters }}
\label{fig:ExpPolar}
\end{figure}
\clearpage

\begin{figure}
\centering
\epsfig{file=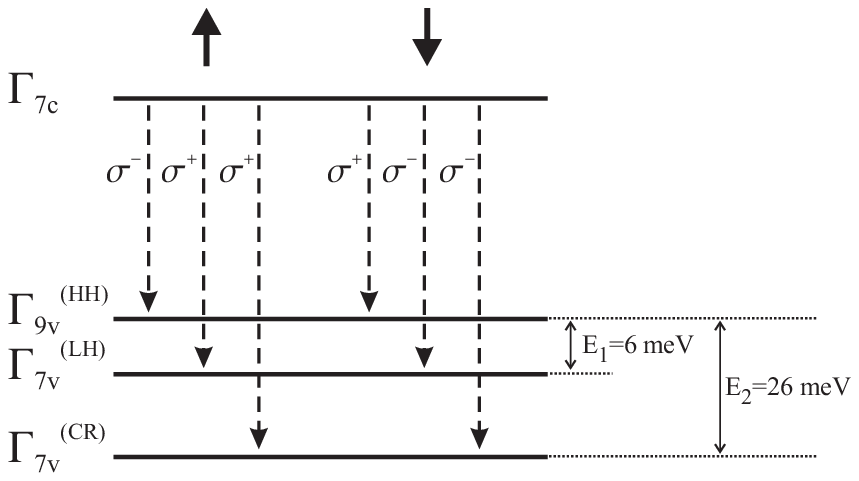,width=3.5in,clip=}
\vspace*{\fill}
\caption{\large \bf C. J. Hill,  Physical Review Letters }{}
\label{fig:bandplot}
\end{figure}
\clearpage

\end{document}